\documentclass[twocolumn,showpacs,preprintnumbers,amsmath,amssymb]{revtex4}

\usepackage{graphicx}
\usepackage{dcolumn}
\usepackage{bm}
\usepackage{epsfig}
\usepackage{subeqn}
\usepackage{epsfig}
\usepackage{amsmath}
\usepackage{amsfonts}

\newcommand{\ba}{\begin{eqnarray}}
\newcommand{\ea}{\end{eqnarray}}
\newcommand{\be}{\begin{equation}}
\newcommand{\ee}{\end{equation}}
\newcommand{\bea}{\begin{eqnarray}}
\newcommand{\eea}{\end{eqnarray}}

\usepackage{theorem}

\theorembodyfont{\rmfamily}

\theoremstyle{break}

\def\QED{~\rule[-1pt]{5pt}{5pt}\par\medskip}

\def\n{\noindent}

\begin{document}


\title{ Geodesics for Efficient Creation and Propagation of Order along Ising Spin Chains}

\author{$^1$Haidong Yuan\footnotemark[1],$^2$Steffen J. Glaser, $^3$Navin Khaneja}

\email{1) haidong@mit.edu, 3) navin@eecs.harvard.edu}

\affiliation{ $^1$Department of Mechanical Engineering, MIT,Cambridge, MA 02139\\
$^2$ Department of Chemistry, Technische Universit\"at M\"unchen, Lichtenbergstr. 4, 85747 Garching \\
$^3$Division of Engineering and Applied Science, Harvard
University, 33 Oxford Street, Cambridge MA 02138 }
\date{\today}

\begin{abstract}
Experiments in coherent nuclear and electron magnetic resonance, 
and optical spectroscopy correspond to control of quantum mechanical ensembles, 
guiding them from initial to final target states by unitary transformations. The 
control inputs (pulse sequences) that accomplish these unitary transformations should 
take as little time as possible so as to minimize the effects of relaxation and 
decoherence and to optimize the sensitivity of the experiments. Here we give 
efficient syntheses of various unitary transformations on Ising spin chains 
of arbitrary length. The efficient realization of the unitary transformations 
presented here is obtained by computing geodesics on a sphere under a 
special metric. We show that contrary to the conventional belief, 
it is possible to propagate a spin order along an Ising spin chain 
with coupling strength $J$ (in units of Hz), significantly faster than $(2J)^{-1}$ per step.   
The methods presented here are expected to be useful for 
immediate and future applications involving control of spin dynamics 
in coherent spectroscopy and quantum information processing.

\end{abstract}


\maketitle
\section{\label{sec:introduction}Introduction}

According to the postulates of quantum mechanics, the evolution of the state 
of a closed quantum system is unitary and is governed by the Schr{\"o}edinger equation. This 
evolution of the state of a quantum system can be controlled by changing the 
Hamiltonian of the system. The 
control of quantum systems has important applications in physics 
and chemistry. In particular, the ability to steer the state of a 
quantum system (or an ensemble of quantum 
systems) from a given initial state to a desired target state forms 
the basis of spectroscopic
techniques such as nuclear magnetic resonance (NMR) and electron spin 
resonance (ESR)
spectroscopy \cite{Ernst, electron} and laser coherent control
\cite{optics} and quantum computing 
\cite{QC1, QC2}. Developing a specific set of control laws - pulse 
sequences- which produce a desired unitary evolution of the state has been 
a major thrust in NMR spectroscopy \cite{Ernst}. For example, in the 
NMR spectroscopy of proteins \cite{Cavanagh}, the transfer of 
coherence along spin chains is an essential step in a large number of 
key experiments. Spin chain topologies have also been proposed as 
architectures for quantum information processing \cite{kane, yamamoto}. In 
practice, the transfer time should be as short as possible, in order to 
reduce the loss due to relaxation or decoherence.      

The time-optimal synthesis of unitary operators
is now well understood for coupled two-spin systems 
\cite{navintoc, Bennett, Timo2spin, General2spin, yuan}.
This problem has also been recently studied in the context of 
linear three-spin topologies 
\cite{navingeodes, navingeodes_exp, navingeodes2}, where
significant savings in the implementation time of trilinear 
Hamiltonians and synthesis of 
couplings between indirectly coupled qubits were demonstrated over 
conventional methods.
In  \cite{navingeodes, navingeodes_exp, navingeodes2}, it was 
shown that the time-optimal 
synthesis of indirect couplings and of trilinear Hamiltonians from linear 
Ising couplings can be 
reduced to the problem of computing singular geodesics. In this paper we extend these 
methods to linear Ising spin chains of size greater than $3$. In
\cite{navin.chain}, the problem of efficient propagation of spin coherences was studied. 
It was shown that by encoding the inital state of a spin into certain 
so called soliton operators, it is possible to propagate the unknown spin 
state through a spin chain at a rate of  $(2J)^{-1}$ per step.
In this paper, we 
extend these ideas to finding efficient ways 
to synthesize multiple-spin order and propagate coherences in 
linear Ising spin chains. Our methods are based on reducing the problem of 
efficient synthesis of the unitary propagator 
with the available Hamiltonians in minimum time to finding 
geodesics on a sphere under a suitable metric detailed subsequently.  
The main result of this paper is that contrary to common belief, 
it is possible to propagate a spin order along an Ising spin chain 
with coupling strength of $J$ (in units of Hz) between nearest neighbours, significantly 
faster than $(2J)^{-1}$ per step.
We present explicit pulse sequences for advancing a spin order along 
an Ising spin at a rate $\sim \frac{\sqrt{2}-1}{J}$, which reduces 
the propagation time to $81\%$ of the best known methods.

\n Numerous approaches have been proposed and 
are currently used \cite{Cavanagh} to transfer
polarization or coherence through chains of coupled spins. 
Examples are pulse trains that create an effective 'XY' Hamiltonian \cite{XYa, XYb, Advances} 
which make it possible to
propagate spin waves in such chains 
\cite{spinwave1, spinwave2, Brueschweiler}.
In order to achieve the maximum possible
transfer amplitude, many other approaches have been developed, that 
rely either on a series of selective transfer steps between adjacent spins 
or on simple concatenations of two such selective transfer 
steps \cite{Cavanagh, Advances}. In our recent work, we showed 
by suitable encodings of the spin operators it is possible to transfer
the unknown state of a spin at the rate of $\frac{1}{2J}$ time 
units per step through the spin chain \cite{navingeodes2}. Subsequently,
the problem has been studied in other contexts, see \cite{jones1, twamley}
and references in there. We show that methods presented here 
are significantly shorter than these state of the art techniques. 
Furthermore, the transfer mechanisms that result from the methods 
presented have a clear geometric interpretation.

\section{\label{sec:theory}Theory}

We consider a linear chain of $n$ spins, placed in a static
external magnetic field in the $z$ direction, with equal Ising type
couplings between next neighbors\cite{Ising1925, Caspers}. In a
suitably chosen (multiple) rotating frame, which rotates with each
spin at its resonant frequency, the free evolution of the spin system 
is given by the coupling Hamiltonian
$$H_c=2\pi J\sum_{m=1}^{n-1} I_{mz}I_{(m+1)z}. $$
We assume that the Larmor frequency of the spins are well separated 
compared to the coupling constant $J$, so that 
we can selectively rotate each spin at a rate much faster than the 
evolution of the couplings. The goal of the pulse
designer is to make an appropriate choice of the control variables
like the amplitude and phase of the external RF field to
effect a net desired unitary evolution.

We first consider the problem of synthesizing a unitary
$U$, which efficiently transfers a single spin coherence
represented by the initial operator $I_{1x}$ to a 
multiple spin order represented by the operator 
$2^{n-1}\prod_{m=1}^{n}I_{mz}$. This transfer is routinely 
performed in multidimensional high resolution NMR 
experiments \cite{Cavanagh}. The techniques developed in 
solving this problem time optimally will be used 
subsequently for other key transfers.

The conventional strategy for achieving this transfer is achieved through 
the following stages
\begin{eqnarray}
\label{eq:H}
\aligned &I_{1x}\stackrel{H_c}\rightarrow2I_{1y}I_{2z}\stackrel{I_{2y}}\rightarrow
2I_{1y}I_{2x}\stackrel{H_c}\rightarrow 4I_{1y}I_{2y}I_{3z}\\
&\stackrel{I_{3y}}\rightarrow4I_{1y}I_{2y}I_{3x}\rightarrow
\cdots\rightarrow 2^{n-1}(\prod_{m=1}^{n-1}I_{my})I_{nz}
\endaligned
\end{eqnarray}
In the first stage of the transfer, operator $I_{1x}$ evolves to
$2I_{1y}I_{2z}$ under the natural coupling Hamiltonian 
$H_c$ in $\frac{1}{2J}$ units of time. This operator is then
rotated to $2I_{1y}I_{2x}$ by applying a hard $(\frac{\pi}{2})_y$ pulse 
on the second spin, which evolves to $4I_{1y}I_{2y}I_{3z}$ under the 
natural coupling, followed by a hard  $(\frac{\pi}{2})_y$ pulse on spin $3$ and so on. 
The final state $2^{n-1}(\prod_{m=1}^{n-1}I_{my})I_{nz}$ prepared in this manner 
is locally equivalent to $2^{n-1}\prod_{m=1}^{n}I_{mz}$. The whole transfer involves
$n-1$ evolution steps under the natural Hamiltonian, each taking $\frac{1}{2J}$
units of time, resulting in a total time of $\frac{n-1}{2J}$.
We now formulate the problem of this transfer as a problem of optimal
control and derive time efficient strategies for achieving this transfer.

\begin{figure}[h]
\begin{center}
\includegraphics[scale=.5]{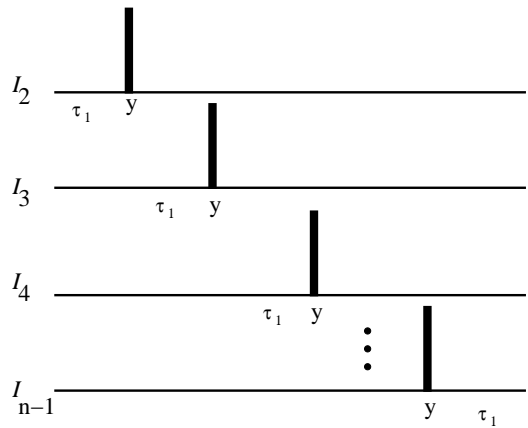}
\end{center}
\caption{ The figure depicts the conventional 
pulse sequence for the creation of multiple spin order as in Equation (\ref{eq:H}).
The spins are sequentially irradiated with hard $(\frac{\pi}{2})_y$ 
pulses with a period of $\tau = \frac{1}{2J}$ between them for the evolution 
of natural coupling.}
\label{fig:2}
\end{figure}

\section{The Optimal Control Problem}

To simplify notation, we introduce the following symbols for the
expectation values of operators that play a key part in the transfer.
\begin{eqnarray}
\label{eq:H1}
\aligned
&x_1=<I_{1x}>\\
&x_2=<2I_{1y}I_{2z}>\\
&x_3=<2I_{1y}I_{2x}>\\
&x_4=<4I_{1y}I_{2y}I_{3z}>\\
&\vdots\\
&x_{2n-3}= <2^{n-1}I_{1y}I_{2y}I_{3y}\cdots I_{(n-1)x}> \\
&x_{2n-2}=<2^{n-1}I_{1y}I_{2y}I_{3y}\cdots I_{(n-1)y}I_{nz}>
\endaligned
\end{eqnarray}

Let $X = (x_1, x_2, x_3, \dots, x_{2n})^T$.
The evolution of the system is
given by
\begin{equation}
\label{eq:transfer} \frac{dX}{dt}= \pi J
    \left(\begin{array}{cccccccc}
      0 & -1 & 0 & 0 & 0 & 0 & \cdots\\
      1 & 0 & -u_1 & 0 & 0 & 0 &\cdots\\
      0 & u_1 & 0 & -1 & 0 & 0 &\cdots\\
      0 & 0 & 1 & 0 & -u_2 & 0 &\cdots\\
      0 & 0 & 0 & u_2 & 0 & -1 &\cdots\\
      0 & 0 & 0 & 0 & 1 & 0   & \cdots\\
      \vdots & \vdots & \vdots & \vdots & \vdots& \vdots & \vdots &
\vdots

      \end{array}\right)X
\end{equation}where
$u_k$ are the control parameters 
representing the amplitude of the 
$y$ pulse on spin $k+1$. 
The problem now is to find the optimal
$u_k(t)$, steering the system from 
$(1,0,0,\cdots,0)$ to $(0,0,\cdots,0,1)$ in 
the minimum time.

In this picture, the conventional transfer method is described as setting 
all $u_k=0$ for $\frac{1}{2J}$ units of time, then 
transfer $x_1=1\rightarrow x_2=1$. Now using the control
$u$, we rotate $x_2=1$ to $x_3 = 1$ in arbitrary small time as control can 
be performed arbitrarily fast (compared to the evolution of couplings).
Now we set $u_k=0$ again, evolve $\frac{1}{2J}$ units of time under natural 
Hamiltonian, transfer $x_3=1$ to $x_4=1$ and so on. This way the total 
time for the transfer is $\frac{(n-1)}{2J}$.

We now show how significantly shorter transfer times are achievable, if we 
relax the constraint that the selective rotations on a given spin are only 
hard pulses. We show that if we let the selective operations be carried out by 
soft shaped pulses, along with evolution of the coupling Hamiltonian, then we can 
achieve significantly shorter transfer time. The analytical 
pulse shapes can be computed by formulating the problem as computation of
geodesics on a sphere under a special metric.
We divide the total transfer into transfer steps. We first transfer
$x_1=1$ to $x_3=x_4=\frac{1}{\sqrt{2}}$, then transfer
$x_3=x_4=\frac{1}{\sqrt{2}}$ to $x_5=x_6=\frac{1}{\sqrt{2}}$ and 
so on and finally at the last transfer $x_{2n-3}=x_{2n-2}=\frac{1}{\sqrt{2}}$ 
to $x_{2n}=1$. The first step is described by the following equation. 
The subsequent steps can also be understood by relating them to the 
following equation (we suppress the factor of $\pi J$ in the front of the equation).
\begin{equation}
\label{eq:transfer2} \frac{d}{dt}\left(\begin{array}{c}
           x_1\\
           x_2\\
           x_3\\
           x_4
           \end{array}\right)=
    \left(\begin{array}{cccc}
      0 & -1 & 0 & 0\\
      1 & 0 & -u & 0\\
      0 & u & 0 & -1\\
      0 & 0 & 1 & 0

      \end{array}\right) \left(\begin{array}{c}
           x_1\\
           x_2\\
           x_3\\
           x_4
           \end{array}\right),
\end{equation}

We make the following change of variables. Let
$$r_1=x_1,r_2=\sqrt{x_2^2+x_3^2},r_3=x_4, \tan \theta=\frac{x_3}{x_2}$$
as depicted in the following picture. 

\begin{figure}[h]
\begin{center}
\includegraphics[scale=.5]{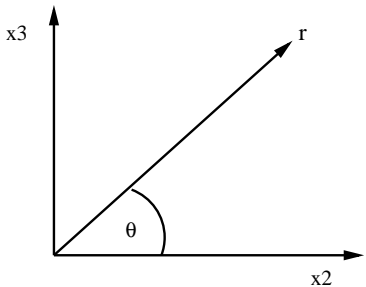}
\end{center}
\caption{ The figure depicts the control angle $\theta$ }
   \label{fig:2}
\end{figure}

Using $u$, we can control the angle $\theta$, so we can think of
$\theta$ as a control variable. The equation for the evolution of
$r_i$ are
\begin{equation}\label{eq:main0}
\frac{d}{dt}\left(\begin{array}{c}
           r_1\\
           r_2\\
           r_3\\
           \end{array}\right)=A\left(\begin{array}{c}
           r_1\\
           r_2\\
           r_3\\
           \end{array}\right),
\end{equation}
where $$A=\left(\begin{array}{ccc}
      0 & -\cos\theta(t) & 0 \\
      \cos\theta(t) & 0 & -\sin \theta(t) \\
      0 & \sin\theta(t) & 0 \\

      \end{array}\right). $$ We now need to find $\theta(t)$ for steering the system from
$(1,0,0)$ to $(0,\frac{1}{\sqrt{2}},\frac{1}{\sqrt{2}} )$ in the minimum time.

We now show that this is equivalent to find the geodesic on the sphere
from $(1,0,0)$ to $(0,\frac{1}{\sqrt{2}},\frac{1}{\sqrt{2}})$
under the metric $g=\frac{dr_1^2+dr_3^2}{r_2^2}$ \cite{navingeodes2}.

The time of transfer $\tau$ can be written as $\int_{0}^\tau
\sqrt{\sin^2 \theta(t) + \cos^2 \theta(t)}\ dt$. Substituting for
$\sin \theta(t)$ and $\cos \theta(t)$ from (\ref{eq:main0}), this
reduces to
$$\int \underbrace{\sqrt{\frac{(\dot r_1)^2 + (\dot
r_3)^2}{r_2^2}}}_{L} dt.$$ Thus minimizing $\tau$ amounts to
computing the geodesic under the metric 
$$ g = {\frac{dr_1^2+dr_3^2}{r_2^2}}. $$ 

The Euler-Lagrange equations for the geodesic take the form
$\frac{d}{dt}(\frac{\partial L}{\partial \dot{r}_1}) =
\frac{\partial L}{\partial r_1}$ and $\frac{d}{dt}(\frac{\partial
L}{\partial \dot{r}_3}) = \frac{\partial L}{\partial r_3}$. Note,
$r_2^2=1-r_1^2-r_3^2$ and along the geodesic, $L$ is constant. We
get \be \label{eq:Hamil}
\frac{d}{dt}\frac{\dot{r}_1}{r_2^2}=L^2\frac{r_1}{r_2^2};
\ \ \frac{d}{dt}\frac{\dot{r}_3}{r_2^2}=L^2\frac{r_3}{r_2^2}
\ee which implies that
$$\frac{d}{dt}\frac{\dot{r}_1r_3-\dot{r}_3r_1}{r_2^2}=0$$
Let $f=\frac{\dot{r}_3r_1 - \dot{r}_1r_3}{r_2^2}$, then $f$ is
constant along the optimal trajectory. For $f$ to
be finite, we need $\sin \theta(0) = 0$, i.e., $\theta(0) = 0$.
From Eq.~\ref{eq:Hamil}, we
get 
\be \frac{d}{dt}(\frac{\dot{r}_1}{r_2})=f\frac{\dot{r}_3}{r_2}
\ee 
Substitute
$\frac{\dot{r}_1}{r_2}=-\cos\theta(t)$,$\frac{\dot{r}_3}{r_2}=\sin\theta(t)$,
gives
$$\frac{d}{dt}(-\cos\theta(t))=f \sin\theta(t),$$
so $\dot{\theta}=f$,
$$\theta(t)=ft+\varphi$$
Since $\theta(0)=0$ we get $\varphi=0$. We can now simply
search numerically for the constant $f$ and 
it turns out that for the transfer from $(1,0,0)$ to
$(0,\frac{1}{\sqrt{2}},\frac{1}{\sqrt{2}})$, the optimal
$f$ (in units of $\pi J$) and the minimum time $\tau_1$ 
(in units of $\frac{1}{\pi J}$) are
\begin{equation}
\label{eq:l1}
f=0.542,\ \ \ \tau_1 = 1.94 \sim \frac{3\pi(\sqrt{2}-1)}{2} 
\end{equation}

\begin{figure}[h]
\begin{center}
\includegraphics[scale=.5]{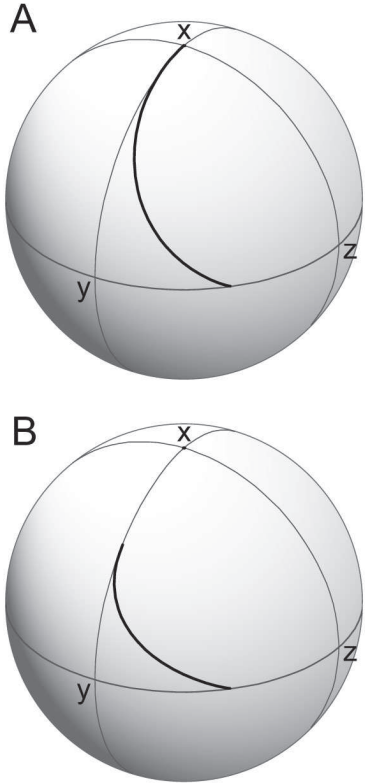}
\end{center}
\caption{ The curve in Fig. A, depicts the shortest path
connecting the north pole $(1, 0, 0)$ to a point $(0, \frac{1}{\sqrt{2}},
\frac{1}{\sqrt{2}})$, under the metric $g$.  The curve in Fig. B, depicts 
the shortest path
connecting $(\frac{1}{\sqrt{2}}, \frac{1}{\sqrt{2}}, 0)$ to a point $(0, \frac{1}{\sqrt{2}},
\frac{1}{\sqrt{2}})$, under the metric $g$. }
   \label{fig:1}
\end{figure} 

\begin{figure}[h]
\begin{center}
\includegraphics[scale=.7]{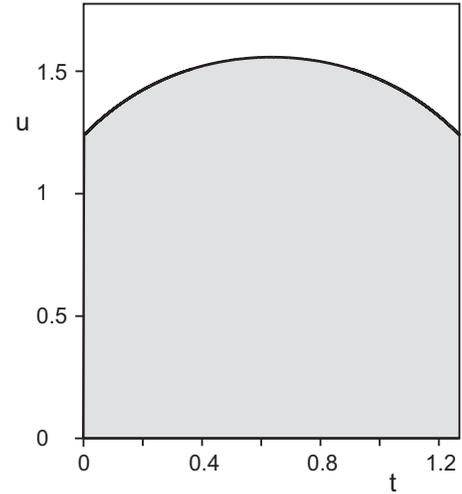}
\end{center}
\caption{\label{fig:4} Fig. shows the optimal $u(t)$ in  
Eq. (\ref{eq:transfer2}) for the transfer
$(\frac{1}{\sqrt{2}}, \frac{1}{\sqrt{2}}, 0, 0)$ to
$(0, 0, \frac{1}{\sqrt{2}}, \frac{1}{\sqrt{2}})$. The time is in the 
units of $(\pi J)^{-1}$. }
\end{figure}

Now differentiating the expression, 
$\frac{x_3}{x_2}(t)= \tan(\theta(t))$ gives that $u(t)$ in is constant 
with a value $u=1.084$.
The subsequent transfer steps, whereby 
$(x_k, x_{k+1})= (\frac{1}{\sqrt{2}}, \frac{1}{\sqrt{2}})$
is transferred to $(x_{k+2}, x_{k+3})= (\frac{1}{\sqrt{2}}, \frac{1}{\sqrt{2}})$,
are described by the same equation of the type as (\ref{eq:transfer2}), except 
we now want to transfer from
initial state $(\frac{1}{\sqrt{2}},\frac{1}{\sqrt{2}},0,0)$ to
final state $(0,0,\frac{1}{\sqrt{2}},\frac{1}{\sqrt{2}})$ for the
intermediate steps and finally from
$(\frac{1}{\sqrt{2}},\frac{1}{\sqrt{2}},0,0)$ to $(0,0,0,1)$
for the last step. For the intermediate steps, using the same method 
as before, we get $f$ (in the units of $\pi J$) and 
the minimum time $\tau_2$ (in units of $(\pi J)^{-1}$) as
\begin{equation}
\label{eq:l2}
f=1.2377, \tau_2 = 1.2692 \sim (\sqrt{2}-1) \pi. 
\end{equation}The optimal $u(t)$ in Eq. (\ref{eq:transfer2}) for this transfer  is
shown in Fig. \ref{fig:4}.
. 

The optimal transfer for the last step 
takes the same time as in (\ref{eq:l1}). The 
total time (in units of $(\pi J)^{-1}$ ) 
is $$ \sim \pi ((n-1)(\sqrt{2}-1)), $$ which is
approximately $81\%$ of the conventional method described earlier.

\begin{figure}[h]
\begin{center}
\includegraphics[scale=.5]{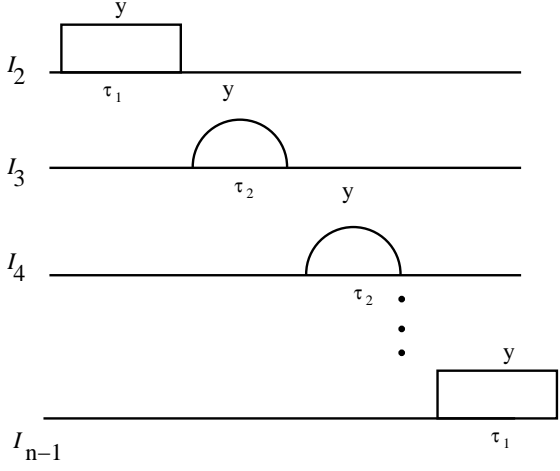}
\end{center}
\caption{ The figure depicts the new pulse sequence 
for the creation of multiple spin order.
The spins $I_3$ to $I_{n-2}$ are sequentially irradiated with shaped pulses as in 
Fig. 4, with pulse duration of $\tau_2$. The couplings 
evolve during the pulse duration and there is no delay between the sequential pulses. The 
initial rectangular pulse of duration
$\tau_1$ on spin 2, creates the state
$(x_3, x_4)= (\frac{1}{\sqrt{2}}, \frac{1}{\sqrt{2}})$ in (\ref{eq:H1}).  
The final rectangular pulse of duration $\tau_1$ on spin $I_{n-1}$, creates 
the desired spin order.}
\label{fig:2}
\end{figure}

We now use these methods for finding efficient ways to 
propagate coherences along Ising spin-chains. 

\subsection{Efficient Transfer of Coherence}

Consider the transfer of coherence 
$$I_{1x} \rightarrow I_{nx}, $$ between the spins $1$ and $n$.

The fastest conventional method for achieving this transfer 
is based on the concatenated INEPT transfer \cite{concatenation}, 
whereby bilinear operators $2 I_{kx}I_{(k+1)z}$ are propagated 
through the spin chain using the following steps:

\begin{eqnarray} 
2 I_{kx}I_{(k+1)z} \stackrel{(H_1)_{\frac{\pi}{2}}}{\rightarrow} -2 I_{kz}I_{(k+1)x} 
 \stackrel{(H_c)_{\frac{\pi}{2}}}{\rightarrow} 2 I_{(k+1)x}I_{(k+2)z}.
\end{eqnarray}

The control Hamiltonian $H_1 = (I_{ky} + I_{(k+1)y})$ evolves for a 
time corresponding to $\frac{\pi}{2}$ rotation, (this step takes arbitrarily small time), 
followed by evolution of the coupling Hamiltonian 
$H_c$ for $\frac{1}{2J}$ units of time. Under this transfer strategy, bilinear 
operators are propagated through the spin chain at a propagation speed of 
$\frac{1}{2J}$. We now show how the propagation rate of coherences through the 
chain can be enhanced by preparing suitable spin states and efficient transfer
of these states using methods described before. Using these ideas, it becomes 
possible to transfer the operators one step along the chain in only 
$\frac{\sqrt{2}-1}{J}$ units of time.

\begin{table}
\begin{center}
\begin{tabular}{|c|c|}
\hline
$\Lambda_{1k}$ & $4I_{kz}I_{(k+1)y}I_{(k+2)x}$ \\
\hline
$\Lambda_{2k}$ & $8I_{kz}I_{(k+1)y}I_{(k+2)y}I_{(k+3)z}$\\
\hline
$\Lambda_{3k}$ & $2I_{(k+1)x}I_{(k+2)x}$\\
\hline 
$ \Lambda_{4k}$ & $4I_{(k+1)x}I_{(k+2)y}I_{(k+3)z}$\\
\hline
$D_{1k}$ & $8I_{kz}I_{(k+1)y}I_{(k+2)y}I_{(k+3)x}$\\
\hline
$D_{2k}$ & $16 I_{kz}I_{(k+1)y}I_{(k+2)y}I_{(k+3)y}I_{(k+4)z}$\\
\hline
$D_{3k}$ & $4I_{(k+1)x}I_{(k+2)y}I_{(k+3)x}$\\
\hline
$D_{4k}$ & $8I_{(k+1)x}I_{(k+2)y}I_{(k+3)y}I_{(k+4)z}$\\
\hline
\end{tabular}
\end{center}
\caption{Relevant operators for transfer of spin order.}
\end{table}
Consider the operator

\begin{eqnarray}
\label{eq:lambda}
\Lambda_{k} &=& \frac{1}{2} ( \Lambda_{1k}  + \Lambda_{2k} + \Lambda_{3k} + \Lambda_{4k} ),
\end{eqnarray}where various operators in the equation are displayed in Table I.
By application of the Hamiltonian $H = H_c + u(t)H_2$, the operator 
$\Lambda_{k}$ is advanced one step in the spin chain, 
$$ \Lambda_{k} \stackrel{H }{\longrightarrow} \Lambda_{k+1}, $$ where 
$H_2 = \pi J (I_{(k+1)y} + I_{(k+3)y})$ and $u(t)$ is the optimally shaped 
pulse that transfers $(\frac{1}{\sqrt{2}},\frac{1}{\sqrt{2}}, 0, 0)$ to  
$(0, 0, \frac{1}{\sqrt{2}},\frac{1}{\sqrt{2}})$ in equation (\ref{eq:transfer2})
in the minimum time. The time of this transfer is $\tau_2 \sim \frac{(\sqrt{2}-1)}{J}$ and it 
advances the spin operators one step in the spin chain. 
This is best understood by splitting the Hamiltonian $H$ into 
two commuting Hamiltonians 
$H_{a} = \pi J (2I_{(k+2)z}I_{(k+3)z} + 2I_{(k+3)z}I_{(k+4)z} + u(t) I_{(k+3)y})$
and $ H_{b} = \pi J (2I_{kz}I_{(k+1)z} + 2I_{(k+1)z}I_{(k+2)z} + u(t) I_{(k+1)y})$, 
whose action can be analyzed separately. We 
identify $(\langle \Lambda_{1k} \rangle, \langle \Lambda_{2k} \rangle, \langle D_{1k} \rangle, \langle D_{2k} \rangle )$ and $(\langle \Lambda_{3k} \rangle, \langle \Lambda_{4k} \rangle, 
\langle D_{3k} \rangle, \langle D_{4k} \rangle )$ with $(x_1, x_2, x_3, x_4)$. 

Under the action of the Hamiltonian $H_{a}$, the 
evolution of the operators $(x_1, x_2, x_3, x_4)$ is exactly identical to the one in 
equation (\ref{eq:transfer2}). If $u(t)$ is chosen optimally, as described, 
in Fig. 4, we transfer the state $(\langle \Lambda_{1k} \rangle, \langle \Lambda_{2k} \rangle, \langle D_{1k} \rangle, \langle D_{2k} \rangle ) = (\frac{1}{2}, \frac{1}{2}, 0, 0)$ to 
$(0, 0, \frac{1}{2}, \frac{1}{2})$. Similarly 
$(\langle \Lambda_{3k} \rangle, \langle \Lambda_{4k} \rangle, 
\langle D_{3k} \rangle, \langle D_{4k} \rangle ) = (\frac{1}{2}, \frac{1}{2}, 0, 0)$ is transferred
to  $(0, 0, \frac{1}{2}, \frac{1}{2})$.
So under $H_a$, $\Lambda_k$ is transferred to $\frac{1}{2} ( D_{1k}  + D_{2k} + D_{3k} + D_{4k} )$.


Similar analysis shows that under the action of $H_{b}$, we transfer
$(\langle D_{1k} \rangle, \langle D_{2k} \rangle, \langle D_{3k} \rangle, 
\langle D_{4k} \rangle ) = \frac{1}{2} (1, 1, 1, 1)$ to 
$(\langle \Lambda_{1(k+1)} \rangle, \langle \Lambda_{2(k+1)} \rangle, 
\langle \Lambda_{3(k+1)} \rangle, \langle \Lambda_{4(k+1)} \rangle ) = \frac{1}{2} (1, 1, 1, 1)$, 
which is the desired operator $\Lambda_{k+1}$.

Creation of the operator $\Lambda_1$ from the operator 
$I_{1x}$ can also be achieved in many ways. Here we 
present one such preparation, (which is not necessarily 
time optimal). Similarly, the final 
operator $\Lambda_{n-3}$ can be collapsed on to the final 
state $I_{nx}$ using steps as before.
\begin{eqnarray}
\aligned &I_{1x}\stackrel{(H_c)_{\frac{\pi}{2}}}\rightarrow2I_{1y}I_{2z}
\stackrel{(I_{2y})_{\frac{\pi}{2}}}\rightarrow
2I_{1y}I_{2x}\stackrel{(H_c)_{\frac{\pi}{2}}}\rightarrow 4I_{1y}I_{2y}I_{3z}\\
&\stackrel{(I_{3y})_{\frac{\pi}{2}}}\rightarrow 
4I_{1y}I_{2y}I_{3x}\stackrel{(H_c)_{\frac{\pi}{4}}}\rightarrow 
\frac{1}{\sqrt{2}} (4I_{1y}I_{2y}I_{3x} + 8I_{1y}I_{2y}I_{3y}I_{4z}).
\endaligned
\end{eqnarray}
Now by applying a $(\frac{\pi}{2})_x$ pulse on the first spin, we will get a 
superposition of $4I_{1z}I_{2y}I_{3x}$ and $8I_{1z}I_{2y}I_{3y}I_{4z}$. Now, 
letting the natural coupling $H_c$ evolve for another 
$\frac{\pi}{4}$ rotation period with spin $4$ is decoupled, followed by 
a $(\pi)_y$ pulse on spin 2, we will generate the operator $\Lambda_1$. 

We now show how these ideas can be generalized for transferring
the unknown state of a spin $1$ to spin $n$. All we need to show
is that we can simultaneously transfer $I_{1x} \rightarrow I_{nx}$
and $I_{1y} \rightarrow I_{ny}$, and this ensures that $I_{1z}$ is
transferred to $I_{nz}$. Furthermore, this ensures that an
arbitrary unknown state of spin $1$, ($aI_{1z}+bI_{1x} + cI_{1y}$),
is transferred to the final state ($aI_{nz}+bI_{nx} + cI_{ny}$).
Consider the pair of operators $(\Lambda_{k}, \Lambda_{k+2})$,
where $\Lambda_k$ is defined in Eq.(\ref{eq:lambda}). Then
observe, $$(\Lambda_{k}, \Lambda_{k+2}) \stackrel{ (H)
}{\longrightarrow} (\Lambda_{k+1}, \Lambda_{k+3}), $$ where
$H=H_c+u(t)\pi J (I_{(k+1)y} + I_{(k+3)y} +  I_{(k+5)y})$ and
$u(t)$ is the same optimal pulse shape as in Eq.(\ref{eq:l2}) and
it takes $\frac{\sqrt{2}-1}{J}$ units of time to advance the pair
$(\Lambda_{k}, \Lambda_{k+2})$, one step in the spin chain. We can
now encode the operators $(I_{1y}, I_{1x})$ via operators
$(\Lambda_{1}, \Lambda_{3})$ and transport it along the chain at a
rate of $\frac{\sqrt{2}-1}{J}$ units of time per step. The
encoding can be achieved by the following steps: first applying a
$(\frac{\pi}{2})_{-x}$ pulse on the first spin, transfer $I_{1y}$ to
$-I_{1z}$, then follow the steps in the last paragraph applying the
pulses transferring $I_{1x}$ to $\Lambda_{1}$, these pulses will
also transfer $-I_{1z}$ to $\frac{1}{\sqrt{2}}
(I_{1y}+2I_{1x}I_{2z})$. We then apply the pulses transferring $$
\Lambda_{1}\stackrel{ (H_1)
}{\longrightarrow}\Lambda_{2}\stackrel{ (H_2)
}{\longrightarrow}\Lambda_{3},$$ where $H_1=H_c+u(t)\pi J (I_{2y}
+ I_{4y}), H_2=H_c+u(t)\pi J (I_{3y} + I_{5y})$ and $u(t)$ is the
optimally shaped pulse that transfers
$(\frac{1}{\sqrt{2}},\frac{1}{\sqrt{2}}, 0, 0)$ to $(0, 0,
\frac{1}{\sqrt{2}},\frac{1}{\sqrt{2}})$ in equation
(\ref{eq:main0}) in the minimum time. These pulses will transfer
$\frac{1}{\sqrt{2}} (I_{1y}+2I_{1x}I_{2z})$ to
\begin{eqnarray}
\aligned &\frac{1}{\sqrt{2}} (I_{1y}+2I_{1x}I_{2z})\stackrel{
(H_1) }{\longrightarrow}\frac{1}{\sqrt{2}}
(2I_{1x}I_{2x}+4I_{1x}I_{2y}I_{3z})\\
&\stackrel{ (H_2) }{\longrightarrow}\frac{1}{\sqrt{2}}
(4I_{1x}I_{2y}I_{3x}+8I_{1x}I_{2y}I_{3y}I_{4z})
\endaligned
\end{eqnarray} then by applying a $(\frac{\pi}{2})_{-y}$ pulse on the first
spin, we will get a superposition of $4I_{1z}I_{2y}I_{3x}$ and
$8I_{1z}I_{2y}I_{3y}I_{4z}$ and let the natural coupling $H_c$,
evolve for another $\frac{\pi}{4}$ rotation period with spin $4$
and $6$ decoupled, followed by a $(\pi)_y$ on spin 2, we
will generate the operator $\Lambda_1$. Thus, we have succeeded in
encoding $(I_{1y}, I_{1x})$ in $(\Lambda_{1}, \Lambda_{3})$.
Compared to the recently developed methods, \cite{navin.chain}
that transport the two operators $(I_{1x}, I_{1y})$ at a rate of
$\frac{1}{2J}$ per step in the spin chain, 
the time taken by the proposed methods is about
$81\%$ of these recently developed techniques.

\section{Conclusion}
In this article, we studied the problem of time optimal creation of 
multiple quantum coherences and transfer of coherences along linear 
Ising spin chains. We showed that by suitably encoding 
the initial state of the spin in certain operators, it is possible to 
propagate these operators along the spin chain at a rate 
$\sim \frac{\sqrt{2}-1}{J}$, which is significantly faster than the 
conventional belief that it takes $\frac{1}{2J}$ to advance spin operators
in a spin chain. The methods developed in this paper may find applications 
in other problems of optimal control of coupled spin dynamics

\end{document}